\documentclass[twocolumn,showpacs,amsmath,amssymb]{revtex4-1}
\usepackage[dvips]{graphicx}
\setkeys{Gin}{width=8.5cm}

\usepackage{amssymb,amsfonts,amsmath}
\usepackage[usenames,dvipsnames]{color}
\usepackage{url}
\usepackage{hyperref}
\usepackage[normalem]{ulem}
\usepackage{cancel}

\begin{document}
\title{Conceptual model of a cognitive system: dynamical system with plastic self-organizing velocity field }
\author{Natalia~B.~Janson}
\email[E-mail: ]{N.B.Janson@lboro.ac.uk}
\author{Christopher~J.~Marsden}
\affiliation{Department of Mathematical Sciences, Loughborough University, Loughborough LE11 3TU, UK}

\begin{abstract}

Spontaneously evolving  living systems can be modelled as continuous-time dynamical systems (DSs), whose evolution rules are determined by their velocity vector fields. We point out that because of their architectural plasticity, biological neural networks belong to a novel type of DSs whose velocity field is plastic, albeit within bounds, and affected by sensory stimuli. 
We introduce DSs with fully plastic velocity fields  self-organising under the influence of stimuli, 
called self-shaping DSs, and propose that a system of this class represents a conceptual model of a cognitive system.
 We propose a simple phenomenological model that within a single field-shaping mechanism carries out a  set of essential cognitive functions 
 without any supervision and online, just like living cognitive systems do.
 The performance of this model is illustrated experimentally with musical examples.  Unlike in artificial neural networks, this mechanism does not produce spurious attractors and in principle does not limit memory size. Implementation of this principle could pave the way  to creating artificial intelligent devices of a new type.

\end{abstract}
 
 
\maketitle

The working principles of a considerable number of man-made devices are based on their ability to evolve spontaneously
 and their suitable models take the form of dynamical systems (DSs) \cite{Andronov_theory_of_oscillations_book}.
 Examples include a pendulum clock \cite{Andronov_theory_of_oscillations_book} and a vacuum-tube circuit producing electromagnetic oscillations \cite{vanderPol_triode_20}.  
 In addition, in the 20th century the dynamical nature of living systems has been appreciated at all levels of their organisation, from a cell to a population of organisms, meaning that their states are continually evolving and thus they can also be modelled as DSs (\cite{MacKey_Glass_clocks}, for a review see \cite{Janson_nonlinear_bio_CP12}). 

\begin{figure}
\includegraphics[width=0.5\textwidth]{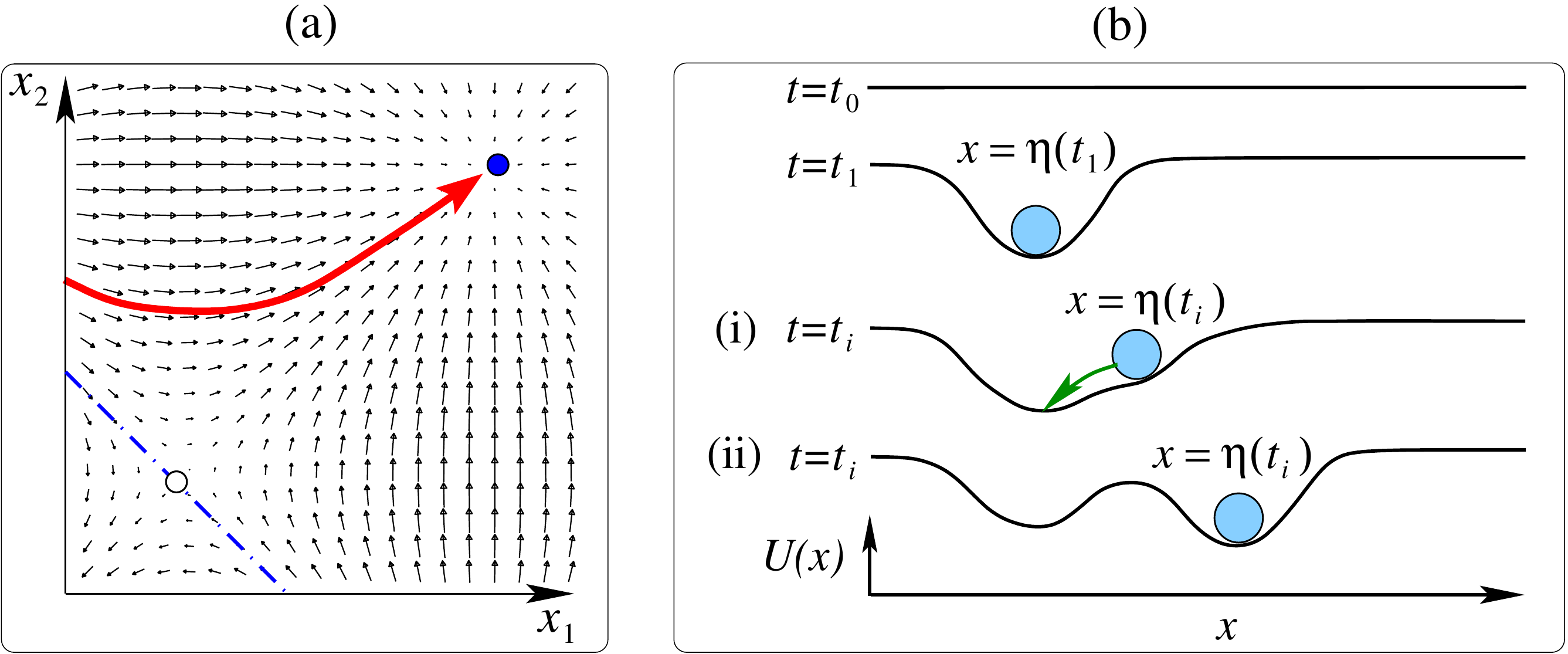}
\caption{\label{fig_vf_ill} (a) For a sample two-dimensional dynamical system the phase trajectory (solid line) is shown being guided by the velocity vector field  (arrows), whose features include a stable fixed point (filled circle) and a manifold outlining the boundary of its basin of attraction (dashed line), and a saddle fixed point (empty circle).
(b) Illustration of the idea of the plastic energy landscape by analogy with memory foam. 
For a one-dimensional  ``foam" stretched in $x$ direction, assume that initially it is flat and described as $U(x,0)$$=$$0$ (see $t$$=$$t_0$). If the quicksilver drop lands onto the ``foam" at position $x$$=$$\eta$, the  
landscape is deformed: a dent appears, which is the deepest exactly at 
$x$$=$$ \eta$, and gets shallower at larger distances from $\eta$ (see $t$$=$$t_1$). Thus the ``foam" will learn about the occurrence of the drop and of its position. 
}
\end{figure}

A DS is a mathematical construct incorporating  a vector $\boldsymbol{x}$$=$$(x_1,\ldots,x_N)$ describing the system state at any time  $t$,  and a rule determining how the state evolves in time. In continuous-time models this 
rule is usually specified by a system of ordinary differential equations 
\begin{equation}
\label{ds}
\frac{\mathrm{d} x_1}{\mathrm{d} t}=s_1(x_1,\ldots,x_N),  \ \ \ldots, \ \  
\frac{\mathrm{d} x_N}{\mathrm{d} t}=s_N(x_1,\ldots,x_N), 
\end{equation}
or in compact vector notation 
\begin{equation}
\label{ds_vec}
\frac{\mathrm{d} \boldsymbol{x}}{\mathrm{d} t}=\boldsymbol{s}(\boldsymbol{x}). 
\end{equation}
Here, $\boldsymbol{s}$$=$$(s_1,\ldots,s_N)$ is the {\it velocity vector field} (Fig.~\ref{fig_vf_ill}(a)), 
which can be loosely understood as a  ``force" that pushes the state $\boldsymbol{x}$ in a certain direction, and is generally different at different points in the phase space. 
This field determines the observable behavior of the system in all feasible situations.

In reality all inanimate and living systems interact with the environment and experience 
time-varying perturbations or stimuli. For their description non-autonomous DSs are usually utilized \cite{Kloeden_nonauton_ds_11}, which differ from (\ref{ds_vec}) in that their velocity fields explicitly depend on time, namely, 
$\frac{\mathrm{d} \boldsymbol{x}}{\mathrm{d} t}=\Tilde{\boldsymbol{v}}(\boldsymbol{x},t). $ 
Since external perturbations are usually describable as time-varying vectors $\boldsymbol{\eta}(t)$, the appropriate models  take the form \cite{Freidlin_random_pert}
\begin{equation}
\label{nonaut_ds_vec_pert}
\frac{\mathrm{d} \boldsymbol{x}}{\mathrm{d} t}=\boldsymbol{v}(\boldsymbol{x},\boldsymbol{\eta}(t)). 
\end{equation}
The performance of such systems can be understood as having every vector of the core velocity field  $\boldsymbol{v}(\boldsymbol{x},\boldsymbol{0})$ (corresponding to the absence of stimulus) forcibly amended at every time $t$. 
Importantly, after the stimulus ceases ($\boldsymbol{\eta}(t)$$=$$0$)  the  field 
$\boldsymbol{v}$ in  (\ref{nonaut_ds_vec_pert}) instantly regains its core form and keeps no memory of influence it might have experienced  previously. While this feature describes most situations well enough, there is one type of systems it does not capture: cognitive systems, including the brain.  Namely, the brain is continually perceiving sensory stimuli and,   remarkably,   
retains information about them 
long after they cease. Moreover, despite sensory stimuli being typically quite random, the brain accumulates information in a consistent and orderly manner. Dynamical hypothesis in cognition suggested that a cognitive system is formed by cognitive agents, represented as DSs,  
and looked quite promising \cite{Smolensky_BBS88,vanGelder_98}. However, the DS theory has not been ready to form the foundation for cognitive science and needed extensions directly relevant to cognitive processing \cite{Crutchfield_BBS98}. The purpose of our paper is to extend the DS theory in  order to provide a dynamical basis for the description of cognition. 

In models of biological neural networks (NNs) individual neurons are usually described as DSs \cite{Izhikevich_DS_neurosci_book}, and inter-neuron connections are incorporated as some functions of neuron states. Thus, the model of the whole NN becomes a DS. 
It is well appreciated that the unique learning ability of the brain is thanks to its architectural plasticity, i.e. to the fact that inter-neuron (synaptic) couplings continually evolve in time \cite{Synaptic_plasticity}. We point out that that the physical plasticity of the brain implies the {\it plasticity of the velocity field} of its DS. In its turn, continual reshaping of the velocity field  implies continual updating of the rules governing collective neural spiking and ultimately the behavior. This observation leads us to propose a conceptual model of a learning system as a DS with plastic velocity field that self-organizes under the influence of stimuli.  
For brevity, we call such systems {\it self-shaping DSs}. We also propose a simple phenomenological model of a cognitive system carrying out a set of essential cognitive functions. 

Although there seems no doubt that in the brain memories are stored in synaptic connections \cite{Bliss_memory_synaptic_couplings_Nat93} and in neural circuits \cite{Liu_memory_neur_circuits_Nat12}, and formed thanks to synaptic plasticity,  the link from the brain architecture to memory representation and learning remains to be properly understood \cite{HBP_report_12}.
An elegant mechanism of memory representation was hypothesized in the theory of {\it artificial} NNs, which are rough phenomenological models of  biological  NNs. Namely, it was proposed that memories could be represented as attractors arising in the phase space of a NN as a result of choosing appropriate values of connections, see review \cite{Grossberg_review_88} and \cite{Hopfield_84}. However, given that NNs are strongly non-linear systems, it is generally impossible to predict or to control where in their phase space the next attractor appears or disappears as a result of the adjustment of connections. Also, the memory capacity of NNs is inevitably bounded \cite{Amit_bounded memory_Netw92} and as a result memories themselves are short-lived \cite{Fusi_memory_Neu05,Fusi_memory_NatNeur07}. Moreover, NNs 
are prone to the uncontrollable formation of spurious attractors not representing any valid memories \cite{Amit_book92}. Finally, while the values taken by the stimuli can in principle be unbounded, the states of the NN are principally bounded by the physical properties of neuron membranes, which makes it even more difficult to relate the incoming patterns with their representation by the NN. 

For our phenomenological model we utilize the idea of attractor coding a memory, but depart from the NN paradigm as a collection of rigid units with bounded states coupled flexibly. Instead, we introduce an idea of a fully plastic spontaneously evolving velocity field {\it directly} amenable to stimulus. In our phenomenological model we allow every stimulus to leave its trace on the velocity field and to contribute to the formation of attractors and their basins, and propose a single field-shaping mechanism that enables the cognitive system to simultaneously memorize and categorize the inputs.

We formulate arguably the simplest  self-shaping DS, 
in which the vector field $\boldsymbol{s}$ is the negative of the gradient $\nabla$ of a scalar energy function $V$  \cite{Hirsch_Smale_gradient_DS_book04}, 
\begin{equation}
\label{eq_particle}
\frac{\mathrm{d} \boldsymbol{x}}{\mathrm{d} t}=-\nabla V(\boldsymbol{x},t)=\boldsymbol{s},
\end{equation}
where $\boldsymbol{x}$ both represents the system state in the $N$-dimensional phase space, and codes a certain pattern.  The state point in (\ref{eq_particle}) behaves like a massless particle placed in a potential energy landscape $V(\boldsymbol{x},t)$ and moves towards the relevant local minimum. Gradient systems can have only attractors of the fixed point type at the minima of $V$.   We assume that $V$ is plastic and undergoes continual reshaping. 

Below we derive an equation describing the shaping of the landscape $V$ in response to the random stimulus. To explain this process, it is helpful to employ 
a loose analogy with the ``memory foam" often used in orthopedic mattresses: this foam takes the shape of a body pressed against it, but slowly regains its original shape after the pressure is removed. We convey the idea using the landscape 
in a one-dimensional space, but it can be extended to the space of any dimension. It helps to use an auxiliary function $U(x,t)$ describing the ``foam" profile depending on a single spatial variable $x$ and time $t$, as illustrated by Fig.~\ref{fig_vf_ill}(b). 
Assume that the landscape is elastic with elasticity factor $k$ that models the capacity of the system to forget. Here, we make a simplified assumption that the deeper the dent at the position $x$ is, the faster the landscape tries to come back to $U$$=$$0$. However, the forgetting term can be modelled in a variety of ways, depending on what the situation requires. 

Let the stimulus act as a sequence of quicksilver drops, which  at consecutive time moments $t_1,t_2,\ldots,t_i,\ldots$ fall on the soft surface at positions 
$x$$=$$\eta(t_i)$. Namely, each drop lands, locally deforms the landscape and slides towards the local minimum while gradually evaporating. Starting from a flat landscape with no features and thus free of memories (Fig.~\ref{fig_vf_ill}(b), $t$$=$$t_0$), the first drop creates a dent centered at $x$$=$$\eta(t_1)$ which becomes the first memory 
 (Fig.~\ref{fig_vf_ill}(b), $t$$=$$t_1$). A subsequent drop lands at a different spot $x$$=$$\eta(t_i)$ and deforms $U$
in one of two ways.  (i) If $\eta(t_i)$ is sufficiently close to an earlier stimulus, the drop amends the respective existing dent 
(Fig.~\ref{fig_vf_ill}(b), (i) $t$$=$$t_i$) and slides to its bottom. Thus $\eta(t_i)$ both amends the existing memory and is  recognised as the one, just like perception of different shades of yellow could form a single memory of a yellow color.  
Alternatively, (ii) if $\eta(t_i)$ is very distinct from any existing memory, it forms its own memory (Fig.~\ref{fig_vf_ill}(b), (ii) $t$$=$$t_i$), and the drop remains at the bottom of the respective dent while evaporating. This process is repeated as more stimuli arrive. This shaping mechanism reminds the kernel density estimation used in statistics \cite{Scott_kernel_92}, however, here it is performed in a continuous time domain under more general assumptions.

\begin{figure}
\includegraphics[width=0.45\textwidth]{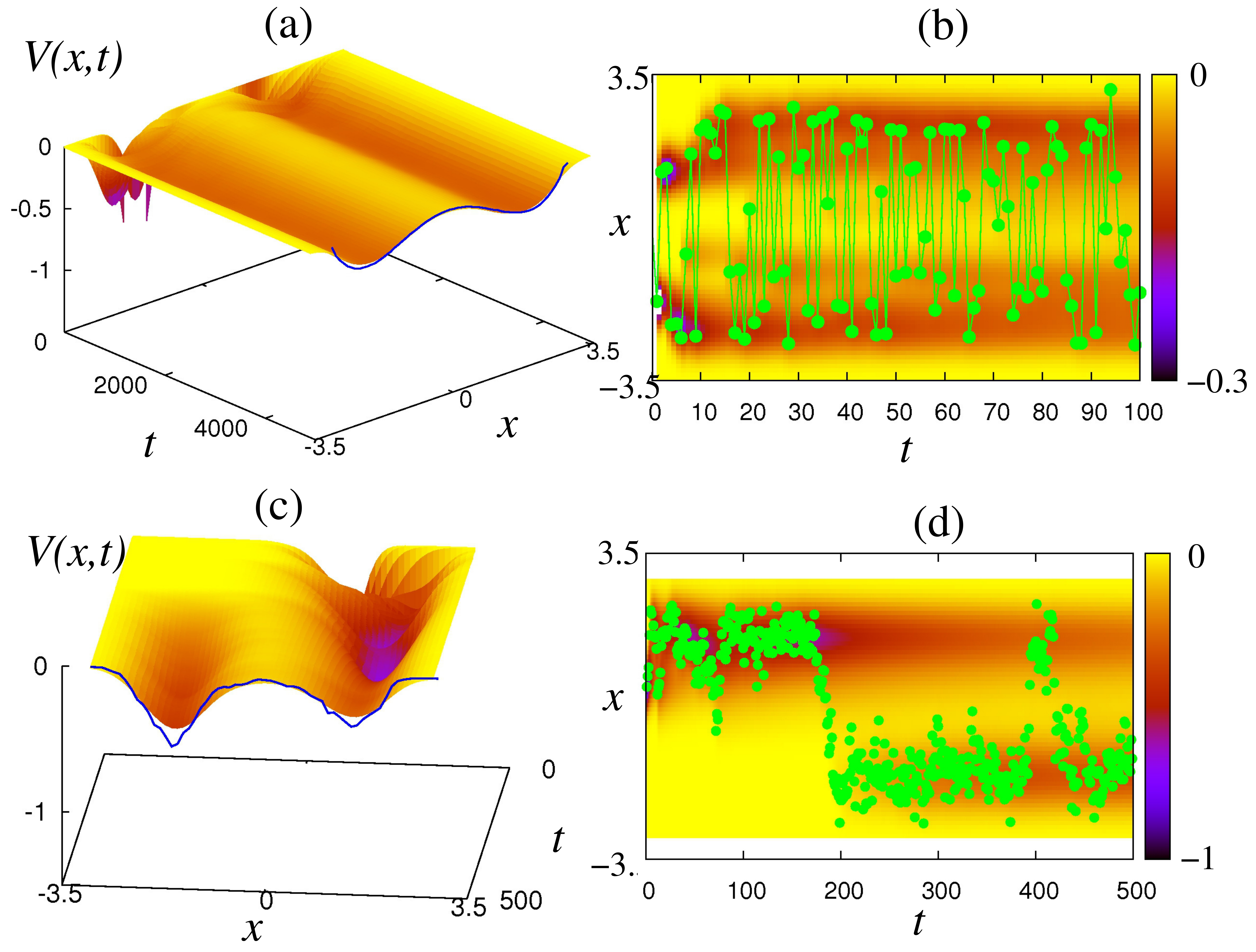}
\caption{\label{fig_uncor_cor} Evolution of the energy landscape $V(x,t)$ as the random stimulus is applied by numerically simulating Eq.~(\ref{eq3}) with $k$$=$$0$: (a,c) 3D view; (b,d) projection of $V(x,t)$ onto $(x,t)$ plane shown by color (shade of grey), and the stimulus applied -- by filled circles. In (a,c) the probability density distribution of stimulus is given by solid line at the front. 
In (a,b) the consecutive values of the stimulus are uncorrelated, and in (c,d) -- correlated. }
\end{figure}

Consider how  $U(x,t)$ changes over a small, but finite time interval $\Delta t$:
\begin{equation}
\label{eq1}
U (x, t +\Delta t) = U (x, t) - g(x - \eta (t)) \Delta t - k U (x, t) \Delta t, 
   \end{equation}
where $g(z)$ is some non-negative bell-shaped function describing the shape of a single dent left by the quicksilver drop, e.g. a  Gaussian function
\begin{equation} 
\label{eq_gauss}
g(z)=\frac{1}{\sqrt{2 \pi \sigma^2_z}} \exp \bigg( - \frac{z^2}{\sigma^2_z} \bigg).
\end{equation}
In (\ref{eq1})  move $U (x, t)$ to the left-hand side, divide both parts by $\Delta t$, and take the
limit as $\Delta t \rightarrow 0$, to obtain 
\begin{equation}
\label{eq2}
  \frac{\partial U (x, t)}{\partial t} = - g(x - \eta (t)) - k U (x, t).
\end{equation}
It can be shown 
that for some arbitrary $\eta (t)$ the solution $U (x, t)$ has
a linear trend and tends to $-\infty$, i.e. it behaves as a linearly decaying function of $t$ with superimposed 
fluctuations. We wish to eliminate this trend and see if we can achieve some sort of stationary behavior of $U (x, t)$. Assuming that $t$$>$$0$ change variables
\begin{eqnarray*}
V = \frac{U}{t}, \quad \frac{\partial V}{\partial t} =  \frac{1}{t} \left(
   \frac{\partial U}{\partial t} - V \right), \quad 
  \frac{\partial U}{\partial t} = t \frac{\partial V}{\partial t} + V, 
\end{eqnarray*}
and rewrite (\ref{eq2}) as follows
\begin{equation}
\label{eq3}
  \frac{\partial V}{\partial t} = - \frac{1}{t} \bigg( V + g(x - \eta(t))  \bigg) - k V.
\end{equation}
In an $N$-dimensional phase space (\ref{eq3}) becomes 
\begin{equation}
\label{eq_self-shaping}
  \frac{\partial V}{\partial t} = - \frac{1}{t} \bigg( V + g(\boldsymbol{x} - \boldsymbol{\eta}(t))  \bigg) - k V,
\end{equation}
where $\boldsymbol{x}$ and $ \boldsymbol{\eta}(t)$ are vectors. 
Within models (\ref{eq3}) and (\ref{eq_self-shaping}) the landscape $V$ and the respective velocity field of Eq.~(\ref{eq_particle}) progressively smooth out and stabilize, as illustrated in Fig.~\ref{fig_uncor_cor} for a one-dimensional system subjected to 
stimulus $\eta(t)$ 
with a two-peak distribution. Equations (\ref{eq_particle}) and (\ref{eq_self-shaping}) form the model of a simple self-shaping DS, in which every value of $\boldsymbol{\eta}$ plays two roles: it both deforms the landscape in (\ref{eq_self-shaping}) and resets initial conditions in (\ref{eq_particle}).

Firstly, we demonstrate the performance of the self-shaping DS using numerically generated stimuli with different statistical properties. 
Fig.~\ref{fig_uncor_cor} shows evolution of $V(x,t)$ as two kinds of scalar stimuli $\eta(t)$ are applied to the one-dimensional  system (\ref{eq3}) \cite{SM_3}. The values of stimuli arriving at consecutive time moments are depicted by filled circles in (b) and (d), and their distributions have similar two-peak shapes shown by the solid lines at the front of (a,c).  The difference between the two inputs is in the amount of temporal correlations in the respective random processes: in (a,b) two consecutive values are uncorrelated  and in (c,d) correlated. 
The stimulus illustrated in Fig.~\ref{fig_uncor_cor} (a,b) is obtained by taking Gaussian white noise and applying a non-linear transformation that changed its distribution into a two-peak one shown in (a) by solid line. This way, the consecutive values of the process remained   uncorrelated. The stimulus in (c,d) is obtained by applying Gaussian white  noise to a differential equation describing a particle moving in a   non-symmetric double-well potential with large viscosity \cite{Malakhov_97}. The distribution of the stimulus is shown in (c) by solid line, and its consecutive values are correlated.

The actual signals applied are shown by filled circles in (b,d), and in $g(z)$ described by (\ref{eq_gauss}) we used $\sigma_z$$=$$\sqrt{0.1}$. One can see that eventually both landscapes shape into the respective distributions of stimuli, but if the stimulus values are uncorrelated, the convergence is faster. If the random process producing the stimulus is not stationary,  $V$ evolves into a {\it time-averaged} distribution of the input. 

\begin{figure}
\includegraphics[width=0.45\textwidth]{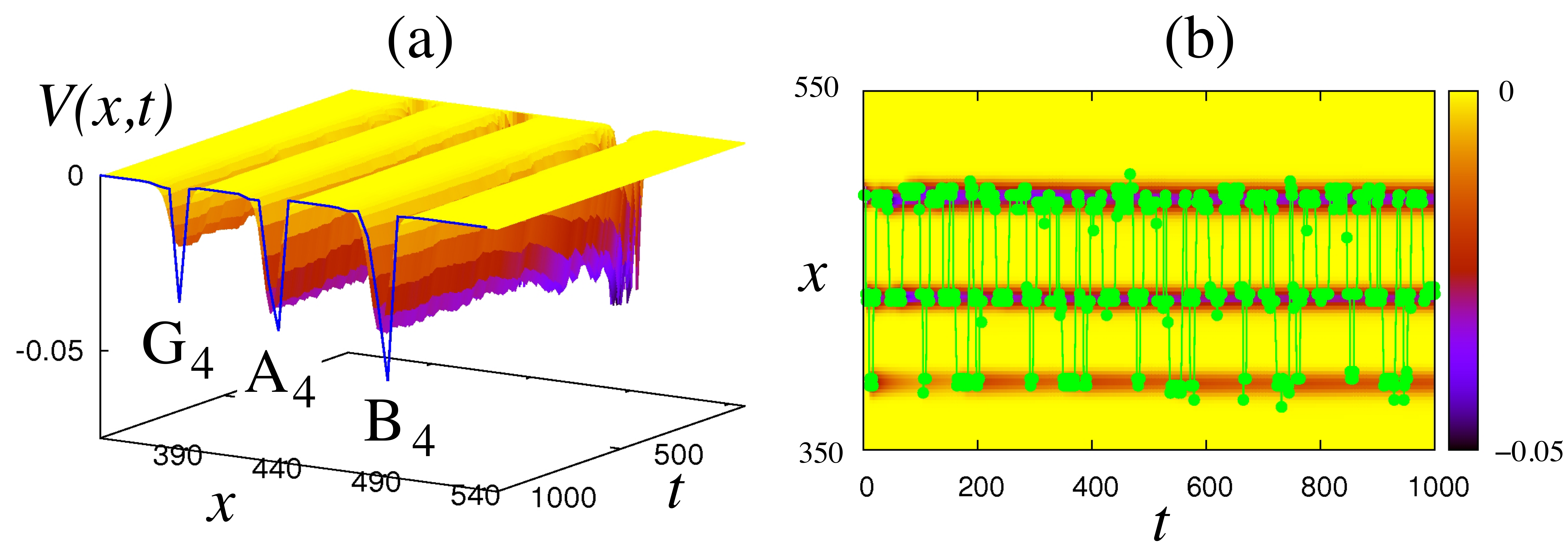}
\caption{\label{fig_flute_1d} (Color online.) Musical note recognition. (a) Evolution of the energy landscape $V(x,t)$ in response to a musical signal performed by an amateur musician. Local minima that develop eventually are very close to the frequencies of the musical notes $G_4$, $A_4$ and $B_4$ that enter the song.  (b)  Filled circles show the actual values of the input, and the shade of the background shows the depth of the energy function.  }
\end{figure}

Next, we illustrate how the proposed gradient self-shaping system automatically discovers and memorises musical notes and phrases. 
A children's song ``Mary had a little lamb" was performed with a flute by an amateur musician six times.  The song involves three musical notes ($A$, $B$ and $G$), consists of 32 beats and was chosen for its simplicity to illustrate the principle. The signal was recorded as a 
wave-file  with sampling rate $8$kHz. In agreement with what is usually done in speech recognition \cite{Flanagan_77}, 
the short-time Fourier Transform was applied \cite{Allen_SFT} to the waveform with a sliding window of duration $\tau$$=$$0.75$ sec, which was roughly equal to the duration of each note. The highest spectral peak was extracted for each window, which corresponded to the main frequency $f$ Hz of the given note. A sequence of frequencies $f(t)$ was used to stimulate the system (\ref{eq3}). 
Note, that each value of $f(t)$ was slightly different from the exact frequency of the respective note, because of the natural variability introduced by a human musician, and the signal $f(t)$ was in fact random, as seen from Fig. \ref{fig_flute_1d}(b).

\begin{figure}
\begin{center}
\includegraphics[width=0.45\textwidth]{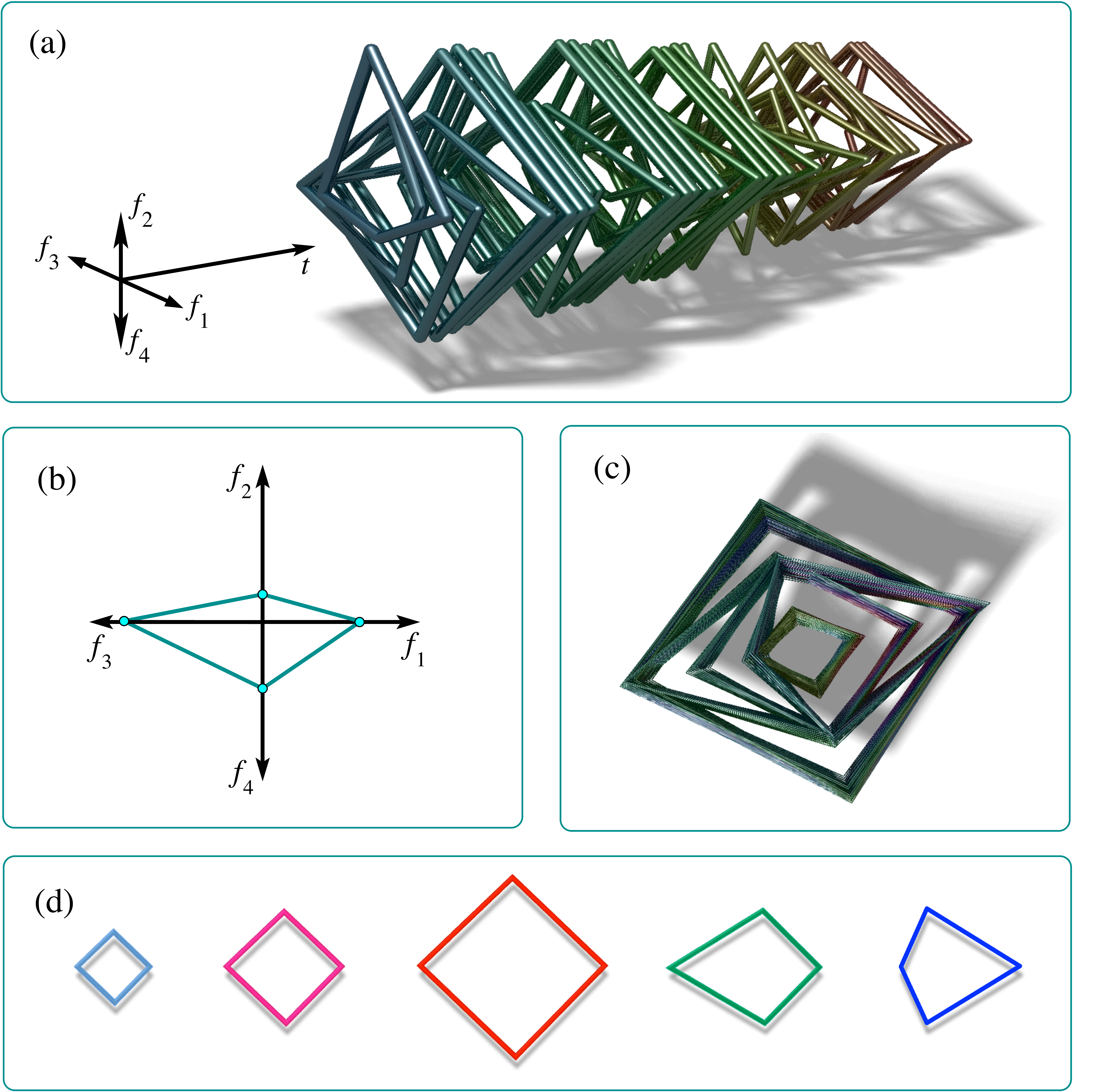}
\caption{\label{fig_4d_music}  (color online) Automatic discovery and memorization of musical phrases. (a) The sight of music: a sequence of melody pieces represented as polygons (as explained in (b)). The shade (color online) is only used to enhance visualization and carries no additional information.   (b) Representing a four-note phrase with a  polygon. (c) The resultant snapshot of the four-dimensional landscape $V(\mathbf{x},t)$. The function $V$ is shown with the smallest values on the top for visualization purposes. The sharp edges on the top represent the most typical musical patterns. (d) The most typical musical phrases automatically detected by the system (\ref{eq_particle}), (\ref{eq_self-shaping}). }
\end{center}
\end{figure}

Firstly, we illustrate how individual musical notes can be automatically identified. A one-dimensional system  (\ref{eq3}) received the signal $\eta(t)$$=$$f(t)$, resampled to $8$Hz to save computation time.  The function $f(t)$ can be seen as a realization of a 1st-order stationary and ergodic  process  $F(t)$, consisting of infinitely many repetitions of the same song, which we observe during finite time. This process has a one-dimensional distribution $p_1^F(f)$, which does not change in time.
A Gaussian kernel $g(z)$ was used with  $\sigma_z$$=$$\sqrt{5}$ Hz. As shown in 
Fig.~\ref{fig_flute_1d}(a), the energy converges to some distribution (with negative sign) shown by the solid line. The most probable frequencies are automatically discovered as the energy minima as follows, with figures in brackets showing the exact frequencies of the respective musical notes: 
434Hz (440Hz) for $A_4$, 490Hz (493.88Hz) for $B_4$, and 388Hz (392Hz) for $G_4$. 

Secondly, we show how system (\ref{eq_self-shaping}) can discover and memorize temporal {\it patterns} -- musical phrases consisting of four beats. The 4D ``foam" was used, and to each of its channels the same signal $f(t)$ was applied, but with a phase shift. Namely, at each time $t$ the system (\ref{eq_self-shaping}) received a vector stimulus $\boldsymbol{\psi}(t)$$=$$(f(t),f(t+\tau),f(t+2\tau),f(t+3\tau))$, $\tau$$=$$0.75$ sec. The procedure of creating a vector with the coordinates made of the delayed versions of the same signal is called delay embedding \cite{Takens_embedding}. For the purpose of this part, we can regard  $\boldsymbol{\psi}(t)$ as a realization of a 4th-order stationary and ergodic vector random process $\boldsymbol{\Psi} (t)$ (which we observe during finite time) with $4$-dimensional  distribution $p^{\boldsymbol{\Psi}}_4(f_1,f_2,f_3,f_4)$. 
We used a multivariate Gaussian kernel $g$ with $\sigma_z$$=$$\sqrt{5}$ Hz in all of its four variables. 

One cannot visualize evolution of a 4D landscape in the same way as we did in Figs.~\ref{fig_uncor_cor}-\ref{fig_flute_1d}, and we
use an alternative representation. We take four half-axes and make their origins coincide (Fig.~\ref{fig_4d_music}(b)). 
For each feasible input $\boldsymbol{\psi}$$=$$(f_1,f_2,f_3,f_4)$ we put 4 points with coordinates $f_i$ on each of half-axes, and connect them by lines. Thus, any feasible input pattern is represented by a polygon on a plane \cite{comment1}. 
The value of $V$ for each value of $\boldsymbol{\psi}$ is depicted in Fig.~\ref{fig_4d_music}(c) as the altitude of the respective polygon, with the deepest polygons shown on top. 
Due to overlapping of polygons it might be difficult to identify the highest ones, which code the most probable musical phrases. However,   one can also find them by numerical simulation using the paradigm of a particle in the 4D landscape that will go to one of the local minima. 
Five most probable polygons are given in smaller scale in Fig.~\ref{fig_4d_music}(d). Recognition of musical phrases is also illustrated with the supplementary  audio files \cite{URL}. 

To conclude, our phenomenological model (\ref{eq_particle}), (\ref{eq_self-shaping})  of a simple cognitive system 
learns  by gradually refining the structure of its velocity field with account of the stream of incoming data. Namely,
in agreement with how humans are creating new memories, and also altering old memories each time they recall them 
 \cite{Schiller_memory_reconsilidation_FBN11}, it creates new attractors (i.e. memorizes new categories), and alters the shapes of the existing attractor basins (i.e. the size and the content of the existing categories) and the attractor locations (i.e. features of the central category members).  
 
 Unlike artificial NNs, the proposed DSs  with fully plastic self-organising velocity field  have no prototype in the physical world. However, 
their operating principle could epitomize the one of both artificial and biological NNs. Namely, we hypothesize that to recognize familiar patterns the NN needs to acquire the velocity field of a certain shape, and to learn continually the NN should be adjusting this shape in response to the new stimuli. In other words, self-organization of the velocity field is the {\it purpose} of the NN,  and  the synaptic plasticity is the {\it means} to achieve this goal. If correct, this hypothesis could give rise to the top-down approach, which is very much needed, but  currently under-represented in brain studies  \cite{Abbott_2013,Arber_2013}. Namely, if one guesses how the velocity field needs to be shaped and amended in a NN forming and recalling memories, one could deduce how synapses and parameters of individual neurons could/should be varying to achieve this, and then verify this deduction experimentally. Thus, it could be possible 
to understand why the laws of synaptic plasticity  need to be this way. 

Unlike artificial NNs, our model can in principle keep an unlimited number of memories because its phase space is unbounded and its velocity field can reshape locally without destroying the global picture. For the same reason our model does not create spurious memories. Implementation of these principles in hardware could pave the way to artificial intelligent devices of a new type. The model can be developed further to give rise to attractors with internal dynamics, 
i.e. limit cycles, quasiperiodic and chaotic attractors, mimicking the performance of networks of ever-spiking biological neurons. 

The authors are grateful to Alexander~Balanov for thorough reading and helpful critical comments on all drafts of this paper, 
and to Victoria~Marsh for playing the flute. 

\newpage

\widetext

~\vspace{0.2cm}

\begin{center}

{\large \bf Supplemental Material\\
 Conceptual model of a cognitive system: dynamical system with plastic self-organizing velocity field} 

~\vspace{-0.2cm}

{ Natalia~B.~Janson, Christopher~J.~Marsden} 

~\vspace{-0.2cm}

\end{center}

\setcounter{equation}{0}
\setcounter{figure}{0}
\setcounter{table}{0}
\setcounter{page}{1}
\makeatletter
\renewcommand{\theequation}{S\arabic{equation}}
\renewcommand{\thefigure}{S\arabic{figure}}
\renewcommand{\bibnumfmt}[1]{[S#1]}
\renewcommand{\citenumfont}[1]{S#1}



The performance of the one-dimensional self-shaping DS (4) and (8) is illustrated in Fig.~2 
 using two numerically simulated examples, and  function $g(z)$ was of the form determined by 
Eq.~(6) with $\sigma_z=\sqrt{0.1}$.  In both examples random signals $\eta(t)$ 
consisted of values from two different categories.  
However, in the first example (Fig.~2 (a,b)) the successive values of input are statistically independent of each other (uncorrelated), while in the second example (Fig.~2 (c,d)) the subsequent values statistically depend on what the previous values were (correlated). 

For the first example we create 
signal $\eta(t)$ by taking Gaussian white noise $\xi(t)$ with zero mean and unit variance and applying a non-linear zero-memory transformation to it, i.e.  $\eta(t)=F(\xi(t))$ \cite{Stratonovich_vol1}. Here $F$ is a non-linear function, chosen in such a way that the probability density of $\eta(t)$ becomes 
$p^{\eta}_1(\eta)$$=$$0.01245 \eta^4 + 0.1065 \eta^2 + 0.0482$ with $\eta \in [-3,3]$, whose negative is shown in 
Fig.~2(a) by solid line at the front. The two peaks in the density describe two categories of input values, and the peak tops (well bottoms in Fig.~2(a)) represent the most typical values from each category. 

In the resultant signal $\eta(t)$, whose portion is shown in  
Fig.~2(b) by filled circles, the consecutive values are {\it uncorrelated}. 
Evolution of the landscape $V(x,t)$ is illustrated in Fig.~2(a) by a surface, and one can see that with time it converges to the density $p^{\eta}_1(x)$ taken with negative sign. 
Another illustration of the process of shaping is given in Fig.~2(b) by the shades (color online) of the background, where the darker shade (color) represents a deeper landscape. The landscape becomes instantaneously deeper at the spot where the new value of 
$\eta$ appears, but progressively smoothes out. 

In the second example, $\eta(t)$ is the numerical solution of the following stochastic differential equation
\begin{equation}
\label{eq_uncor}
\frac{\mathrm{d} \eta}{\mathrm{d} t}=h(\eta)+0.5 \xi(t),
\end{equation}
where $\xi(t)$ is Gaussian white noise with zero mean and unit variance, and $h(\eta)$$=$$3 (\eta-\eta^3)/5$. 
Eq.~(\ref{eq_uncor}) describes a particle moving in a  double-well potential shaped as the negative of an integral of $h(\eta)$, under the assumption of a large viscosity, under the action of a stochastic force. The consecutive values of 
$\eta(t)$ are correlated [24] as illustrated in Fig.~2(d) by filled circles:  if at a certain time the input value is from one category, at the next time moment the input value is more likely to be from the same category, and the switches between different categories occur rarely.  The negative of the density $p^{\eta}_1(x)$, estimated numerically from the realisation of $\eta(t)$ as a distribution histogram, is shown by a solid line in 
Fig.~2(c), and one can see that the landscape $V(x,t)$ approximately tends to this function as time goes by. The shaping of the same landscape is also illustrated in Fig.~2(d), where the depth of $V$ is shown against the values of stimulus $\eta$. 

 One can see that eventually both landscapes in Fig.~2 shape into the negatives of the respective densities, but if the stimulus values are uncorrelated, the convergence is faster.


\begin{thebibliography}{}
\bibitem{comment1} The same method can be used to represent on a plane a vector of any dimension.
\end{thebibliography}
\end{document}